\documentclass[11pt]{article}
\usepackage{amsmath,amsfonts,amssymb,latexsym}

\numberwithin{equation}{section}
\def\be{\begin{equation}}
\def\ee{\end{equation}}
\def\bq{\begin{eqnarray}}
\def\eq{\end{eqnarray}}
\def\beq{\begin{eqnarray*}}
\def\eeq{\end{eqnarray*}}

\def\a{\alpha}
\def\b{\beta}
\def\g{\gamma}

\def\d{\delta}

\def\z{\zeta}

\def\t{\theta}
\def\s{\sigma}

\newcommand{\GD}{\delta}

\newcommand{\GT}{\tau}

\begin{document}
\begin{titlepage}
\title{\huge{The initial state of generalized radiation universes}}
\author{\Large{Spiros Cotsakis\footnote{\texttt{email:\thinspace skot@aegean.gr}},
Georgios Kolionis\footnote{\texttt{email:\thinspace gkolionis@aegean.gr}}\,
and Antonios Tsokaros\footnote{\texttt{email:\thinspace atsok@aegean.gr}}} \\
{Research Group of Geometry, Dynamical Systems and
Cosmology}\\
{University of the Aegean}\\
{Karlovassi 83 200, Samos, Greece}}
\maketitle
\begin{abstract}
\noindent We use  asymptotic methods to study the early time stability of isotropic and homogeneous solutions  filled with radiation which are close initially to the exact, flat, radiation solution in quadratic lagrangian theories of gravity. For such models, we analyze all possible modes of approach  to the initial singularity and prove the essential uniqueness and stability of the resulting asymptotic scheme in all cases except perhaps that of the conformally invariant Bach-Weyl gravity. We also provide a formal series representation valid near the initial singularity of the general solution of these models and show that this is dominated at early times by a form in which both  curvature and radiation play a subdominant role. We also discuss the implications of these results for the generic initial state of the theory.
\end{abstract}
\thispagestyle{empty}
\end{titlepage}

\section{Introduction}      
Cosmology in higher order  gravity theories has become a field of special importance for a number of basic questions about the structure of the early universe.  In particular, stability issues for cosmological solutions in higher order gravity theories currently form a subfield of generalized cosmological theory, one that combines interesting mathematical problems with the search for  fundamental physical effects that might have taken place during the earliest moments of the universe. The cosmological stability problem has two main facets, the perturbation theory aspect used mainly in studies of structure formation, and the asymptotic stability aspect, basic for nonlinear and geometric dynamics issues. The latter problem naturally emerges when one is interested to see whether or not a given exact solution of the field equations has possibly a more general significance, whether a certain \emph{set} of solutions have particular properties, or whether one needs to decide what can happen to the fate of any particular universe in this context if one waits long enough.

The problem of asymptotic stability of isotropic and homogeneous cosmological solutions in higher order gravity theories has itself two major aspects, \emph{late-time stability}-that is deciding the fate of these universes in the distant future, and \emph{early-time stability}-examining the past evolution at early times towards a possible initial singularity. Renewed interest in late time evolution cosmological problems in higher order gravity emerged after the seminal work of Barrow and Ottewill \cite{ba-ot} where the issue of existence and stability of various cosmological solutions with an emphasis on the de Sitter and FRW ones was studied (see also \cite{co-fl93} for related general late-time stability results for Friedman universes in the same context). Two important late-time stability problems that have received a lot of attention are the \emph{cosmic no-hair conjecture} and the \emph{recollapse problem}. For the original cosmic no-hair theorems in higher order gravity, see \cite{maeda,mij-stein,cou-mad}, for limitations of this property cf. \cite{ba87a,ba87b,ba-her} and for generalizations in higher order gravity of the corresponding situation in general relativistic cosmology cf. \cite{page,m-s-star,s,ber1,ber2,co-fl93b}. For various recollapse theorems in generalized cosmological theory see \cite{co-mir1,co-mir2}, and for more elaborate and complete  approaches and results, cf. \cite{mir1,mir2,mir3}. It is interesting that the two issues, cosmic no-hair and recollapse of closed models are not unrelated, in fact a ``premature'' recollapse problem in closed universes that inflate has been formulated and studied, cf. \cite{ba88rec}.

The early-time evolution problem for isotropic and homogeneous cosmologies in higher order gravity is concerned with understanding the host of different behaviours that may occur near the initial singularity. The duality between bouncing and singular early time solutions was already present in the first papers of the subject, cf. \cite{RR,ker}. These first solutions were shown to have the remarkable property of being horizon-breaking \cite{horizonless1}, but it was later pointed out that they may be unstable \cite{horizonless2}. These first results also led to the realization that the question of the possible \emph{early time asymptotes} of the admissible cosmological solutions of the higher order gravity equations  was more involved \cite{muller}, in fact, it was realized that even the simplest ``radiation fluid'' solution $t^{1/2}$, being a solution in both the radiation filled case \emph{and} in vacuum in these theories, may lead to a completely different set of properties than the situation encountered in general relativity. General properties of stability and conditions for instability of the early time flat and curved, radiation-filled, isotropic solutions  were investigated in \cite{co-fl95}, wherein although the interest then was in finding instability properties of these systems as they are followed towards the initial singularity\footnote{It is well known that the corresponding radiation solutions in general relativity are unstable with respect to any kind of perturbation and are also non generic, cf. \cite{ll}.}, there are various possible \emph{stability} results one may deduce from the general conditions and equations by selecting appropriately various constants to take on specific forms. Stable solutions in vacuum near the initial singularity were obtained in the interesting works \cite{ba-mid1,ba-mid2} when one includes in the basic quadratic lagrangian form $R+\a R^2$ a term of the form $\textrm{Ric}^{2n}, n\in\mathbb{Q}$, and considers a flat isotropic cosmology. The method of these papers was to use a linear perturbation analysis of the $t^{1/2}$ solution in vacuum and for \emph{flat} FRW universes, and show that the various perturbations decay to zero asymptotically at early times. It is also known that this vacuum solution is stable under anisotropic, spatially homogeneous  perturbations, cf. \cite{co-dem-derop-que,bh}. It is an interesting open problem to decide therefore the precise extent that a generic perturbation of the flat, vacuum, $t^{1/2}$ solution occupies in the whole space of solutions of the higher order gravity equations.

Consequently, there are two separate early-time asymptotic problems concerning the stability of the flat $t^{1/2}$ FRW solution in higher order gravity: Its stability as a solution of the vacuum field equations, and that as solution of the higher order gravity field equations filled with a radiation fluid. Both problems  need to be examined at all the various levels of stability: For the flat, radiation solution, we know \cite{K1} that it is stable in the space of all flat solutions of the theory asymptotically at early times in four spacetime dimensions. It remains to examine the precise behaviour of this solution consecutively with respect to curved FRW perturbations, anisotropic perturbations and generic inhomogeneous perturbations. As for the vacuum early time problem, as noted above, for the flat, vacuum solution, there are clear indications that it is stable with respect to various FRW and anisotropic perturbations, the strongest known results being its stability with respect to anisotropic perturbations \cite{bh}, and with respect to perturbations in the $R+\b R^2$ action by adding a term of the form $\textrm{Ric}^{2n}$, cf. \cite{ba-mid1,ba-mid2}.

In this paper, we focus on the problem of the early-time behaviour of the flat radiation $t^{1/2}$ solution of higher order gravity with respect to curved FRW perturbations. That is, considered as a solution of the curved FRW equations for the  $R+\b R^2$ action, what is the behaviour as we approach the initial singularity, i.e., as $t\rightarrow 0$ of all solutions  which are initially (that is, for some $t^*>0$) near this radiation solution? For this purpose, we approach the problem via the use of the \emph{method of asymptotic splittings} developed in \cite{CB,go}, and trace all possible asymptotic behaviours that solutions to the higher order curved FRW equations may develop at early times. Following this geometric approach, we are able to show that the radiation solution is stable asymptotically at early times, meaning that the initial state of these universes proves to be a very simple one indeed. Given that this theory is known to admit an inflationary stage \cite{star}, this also means that any pre-inflationary period in such universes is necessarily  isotropic and flat.

The plan of this paper is as follows. In the next Section, we write down the field equations as a dynamical system in suitable variables thus defining a vector field, the curvature-radiation field.  In Section 3, we find  the possible dominant asymptotic forms which might be assumed by the curvature-radiation field towards the initial singularity, and prove the essential uniqueness of the acceptable asymptotic decompositions. In Section 4, we construct formal series expansions compatible with the dominant balances at the initial state and prove that these developments correspond to a local expansion of the \emph{general} solution around the singularity.  We conclude in Section 5, also commenting on  our current results in the larger picture  of ongoing and possible future work.

\section{The basic curvature-radiation field}   
Below we shall be interested in the nature of the dynamics of FRW universes near the initial singularity. These cosmologies are determined by the Robertson-Walker metric of the form
\be \label{rwmetrics}
g_{4}=-dt^{2}+a^{2}\, g_{3},
\ee
where $a(t)$ denotes the scale factor, while each slice is given the 3-metric
\be
g_{3}=\frac{1}{1-kr^2}dr^{2}+r^2g_{2},
\ee
$k$ being the (constant) curvature normalized to take the three values $0, +1$ or $-1$ for the complete, simply connected,  flat, closed or open space sections respectively, and the 2-dimensional sections are such that
\be
g_{2}=d\theta^{2}+\sin^{2}\theta d\phi^{2}.
\ee
We assume that these spaces are filled with a radiation fluid with energy-momentum tensor
$T_{\mu\nu}=(p+\rho)u_\mu u_\nu +pg_{\mu\nu}$,
where the fluid velocity 4-vector is $u^\mu=\delta^\mu_0$, and we take the equation of state to be $p=\rho/3$.

Our general higher order action is
\footnote{We set $8\pi G=c=1$,  and the sign conventions
 are those of \cite{mtw}.}
\be\mathcal{S}=\frac{1}{2}\int_{\mathcal{M}^4}\mathcal{L}_{\textrm{total}}d\mu_{g},
\ee
where $\mathcal{L}_{\textrm{total}}$ is the lagrangian density of the general
quadratic gravity theory given in the form $\mathcal{L}_{\textrm{total}}=\mathcal{L}(R)+\mathcal{L}_{\textrm{matter}}$, with
\be
\mathcal{L}(R)=R + \b R^2 + \g \textrm{Ric}^2 + \d \textrm{Riem}^2 ,
\label{eq:lagra}
\ee
where $\b,\g,\d$ are constants. Since in four dimensions we have the Gauss-Bonnet  identity,
\be \GD \int_{\mathcal{M}^4}R^2_{GB}d\mu_{g}=0,\quad
R^2_{GB}=R^2 - 4\textrm{Ric}^2 + \textrm{Riem}^2,
\label{eq:gentity}
\ee
in the derivation of the field equations
through a $g$-variation  of the action associated with (\ref{eq:lagra}),
only terms up to $\textrm{Ric}^2$ will matter.
For a homogeneous and isotropic space, we have a second useful identity,
 \be
 \GD \int_{\mathcal{M}^4} (R^2 - 3\textrm{Ric}^2)d\mu_{g}=0 \: ,
\label{eq:isontity}
\ee
which further enables us to include the
contribution of the $\textrm{Ric}^2$ term into the coefficient of
$R^2$, altering only the arbitrary constants. Hence, the field
equations read as follows:
\be
R^{\mu\nu}-\frac{1}{2}g^{\mu\nu}R+
      \frac{\xi}{6} \left[2RR^{\mu\nu}-\frac{1}{2}R^2g^{\mu\nu}-2(g^{\mu\rho}g^{\nu\s}-g^{\mu\nu}g^{\rho\s})\nabla_{\rho}\nabla_{\s}R \right]=T^{\mu\nu},
\label{eq:fe}
\ee
where we have set  $\xi=2(3\b+\g+\delta)$. This naturally splits into
$00$- and $ii$-components ($i=1,2,3$), but only the $00$-component of
(\ref{eq:fe}) will be used below.

Using  the metric (\ref{rwmetrics}), the field equation (\ref{eq:fe}) leads to our basic cosmological equation  in the form
\be
\frac{k+\dot{a}^2}{a^2}+\xi\left[2\: \frac{\dddot{a}\:\dot{a}}{a^2} + 2\:\frac{\ddot{a}\dot{a}^2}{a^3}-\frac{\ddot{a}^2}{a^2} - 3\:
\frac{\dot{a}^4}{a^4} -2k\frac{\dot{a}^2}{a^4} + \frac{k^2}{a^4}\right] = \frac{\zeta^2}{a^4} ,
\label{eq:beq}
\ee
where $\zeta$ is a constant defined by the constraint
\be
\frac{\rho}{3}=\frac{\z^2}{a^4},\quad (\textrm{from}\,\,\nabla_{\mu}T^{\mu 0}=0).
\ee
Setting $x=a$, $y=\dot{a}$ and $z=\ddot{a}$, Eq. (\ref{eq:beq}) can be written as an autonomous dynamical
system of the form
\be\label{basic dynamical system}
\mathbf{\dot{x}}=\mathbf{f}_{\,k,\textsc{RAD}}(\mathbf{x}),\quad \mathbf{x}=(x,y,z),
\ee
that is we have the dynamical system
\begin{eqnarray}
\label{eq:ds}
\dot{x} &=& y,\:\:\:\:\:\nonumber\\ \dot{y} &=& z,\:\:\:\:\:\\
\dot{z}& =& \frac{\z^2-k^2\xi}{2\xi x^2y} + \frac{3y^3}{2x^2} + \frac{z^2}{2y} -\frac{yz}{x} - \frac{y}{2\xi}
-\frac{k}{2\xi y} + \frac{ky}{x^2},\nonumber
\end{eqnarray}
 equivalent to the \emph{curvature-radiation} vector field $\mathbf{f}_{\,k,\textsc{RAD}}:\mathbb{R}^3\rightarrow\mathbb{R}^3:(x,y,z)\mapsto\mathbf{f}_{\,k,\textsc{RAD}}(x,y,z)$ with
\be\label{vf}
\mathbf{f}_{\,k,\textsc{RAD}}(x,y,z)=\left( x,y,\frac{\z^2-k^2\xi}{2\xi x^2y} + \frac{3y^3}{2x^2} + \frac{z^2}{2y} -\frac{yz}{x} - \frac{y}{2\xi}
-\frac{k}{2\xi y} + \frac{ky}{x^2}\right).
\ee
The curvature-radiation field $\mathbf{f}_{\,k,\textsc{RAD}}$, or equivalently the dynamical system (\ref{eq:ds}), combines the effects of curvature and radiation and describes completely the dynamical evolution of any radiation-filled FRW universe in higher order gravity. In the following Section we shall see how this field can split asymptotically and determine all dominant modes that it can develop on approach to the initial singularity in higher order gravity.

\section{Asymptotic splittings of the curvature-radiation field}
We are particularly interested below in the behaviour of the universe described by (\ref{vf}) near the initial singularity, taken at $t=0$\footnote{The position of the initial singularity is really arbitrary, and we could have placed it at any $t_0$ and used the variable $\tau=t-t_0$ instead of $t$.}. Any such initial state will be completely described by giving the possible modes of approach of the various solutions to it. These modes are in turn determined by the behavior of (\ref{vf}) near the initial singularity. How can this behaviour be found? For this purpose, we shall use below the method of \emph{asymptotic splittings}, cf. \cite{CB,go}. According to this method, the vector field $\mathbf{f}_{\,k,\textsc{RAD}}$ is asymptotically decomposed in such a way as to reveal its most important dominant features on approach to the singularity. This leads to a detailed construction of all possible local asymptotic solutions valid in the neighborhood of the finite-time singularity. These provide in turn a most accurate picture of all possible dominant features that the field possesses as it is driven to a blow up (for previous applications of this asymptotic technique to cosmological singularities, apart from \cite{K1}, we refer to  \cite{methodapps}).

We expect that the vector field $\mathbf{f}_{\,k,\textsc{RAD}}$ will show some dominant features as we approach the singularity at $t=0$, and these will correspond to the different, inequivalent ways that it splits in the neighborhood of the blow up. To describe the situation precisely, we need two definitions. We say
that a solution $b(t)$ of the system (\ref{eq:ds}) is \emph{asymptotic} to another solution
$a(t)$ provided that the following two conditions hold (the first is
subdivided):
\begin{enumerate}
\item[(i)] Either $(1)$ $a(t)$ is an exact solution of the system, or
 $(2)$ $a(t)$ is a solution of the system (substitution gives $0=0$) as $t\rightarrow\infty$,
\item [(ii)] $b(t)=a(t)[1+g(t)],\, g(t)\rightarrow 0$, as $t\rightarrow\infty$.
\end{enumerate}
If either of these two conditions is not satisfied, then $b(t)$
cannot be asymptotic to $a(t)$. Secondly,  a solution of the system (\ref{eq:ds}) is called \emph{dominant} near the singularity if, for constants $\mathbf{a}=(\theta, \eta, \rho)\in\mathbb{C}^3$, and $\mathbf{p}=(p, q, r)\in\mathbb{Q}^3$, it is asymptotic to the form
\be
\mathbf{x}(t)=\mathbf{a}t^{\mathbf{p}}=(\theta t^{p}, \eta t^{q}, \rho t^{r}). \label{eq:domisol}
\ee
For any given dominant solution of the system (\ref{eq:ds}) near the singularity, we call the pair $(\mathbf{a},\mathbf{p})$ a \emph{dominant balance} of the vector field $\mathbf{f}_{\,k,\textsc{RAD}}$.

Dominant balances, and the corresponding asymptotic integral curves, characterize vector fields near their blow up singularities\footnote{by a solution with a finite-time singularity we mean one where there is a time at which at least one of its components diverges. We note that the usual dynamical systems analysis through linearization etc is not relevant here, for in that one deals with equilibria, not singularities.}. The vector field $\mathbf{f}_{\,k,\textsc{RAD}}$ itself is decomposed asymptotically into a dominant part and another, subdominant part:
\be\label{general split}
\mathbf{f}_{\,k,\textsc{RAD}}=\mathbf{f}^{(0)}_{\,k,\textsc{RAD}} + \mathbf{f}^{\,(\textrm{sub})}_{\,k,\textsc{RAD}},
\ee
and such a candidate asymptotic splitting (or decomposition) needs to be checked for consistency in various different ways before it is to be admitted as such.
By direct
substitution of the dominant balance forms  in our system (\ref{eq:ds}), we look for the possible
scale invariant solutions of the system\footnote{A vector field
$\mathbf{f}$ is called \emph{scale invariant} if
$\mathbf{f}(\mathbf{a}\GT^{\mathbf{p}})=\GT^{\mathbf{p-1}}
\mathbf{f}(\mathbf{a})$, for a more detailed treatment, cf. \cite{CB}.}
\be\label{basic dom ds}
\dot{\mathbf{x}}=\mathbf{f}^{(0)}_{\,k,\textsc{RAD}}(\mathbf{x}).
\ee
How many admissible asymptotic decompositions does the vector field $\mathbf{f}_{\,k,\textsc{RAD}}$ possess on approach to the initial state at $t=0$? We recall that when $k=0$, that is  we have a flat, radiation-filled  FRW model, the vector field $\mathbf{f}_{\,0,\textsc{RAD}}$ has two admissible asymptotic solutions near the initial singularity, as shown in \cite{K1}: In the first family, all flat, radiation solutions are dominated (or attracted) at early times by the form $a(t)\sim t^{1/2}$, thus proving the stability of this solution in the flat case. There is a second possible asymptotic form near the singularity in the flat case, $a(t)\sim t$, but this contains only two arbitrary constants and hence it corresponds to a \emph{particular} solution of the theory (cf. \cite{K1}).

When $k\neq 0$, and we have the present situation of a radiation-filled, curved family of FRW universes to follow asymptotically near the past singularity, the field $\mathbf{f}_{\,k,\textsc{RAD}}$ has more terms-those that contain $k$ in (\ref{vf})-than in the flat case. Since we, as noted above, already have a precise picture of the asymptotic forms of the flat case, we can now study in isolation the combined effects of curvature and radiation alone asymptotically. A simple combinatorial calculation shows that $\mathbf{f}_{\,k,\textsc{RAD}}$ (or the basic system (\ref{eq:ds})) can decompose precisely in
\be
\left(\begin{array}{c}
8\\
1
\end{array}\right)+\left(\begin{array}{c}
8\\
2
\end{array}\right)+\cdots +\left(\begin{array}{c}
8\\
8
\end{array}\right)=255
\ee
different ways. Each one of these 255 different modes leads to an asymptotic splitting of the form (\ref{general split}), each one of which may contain many possible dominant balances and needs to be checked for admissibility. Any candidate asymptotic splitting of the form (\ref{general split}) will be acceptable in principle provided the candidate subdominant part tends to zero asymptotically, that is it indeed behaves as a subdominant contribution to the dominant asymptotic form the field splits into (cf. (\ref{general split})).

It would be exceedingly cumbersome to list here the detailed results we have for each particular decomposition, so we prefer to describe them in a qualitative way instead. The first qualitative conclusion from the asymptotic decomposition analysis we performed is that the first term in the field $\mathbf{f}_{\,k,\textsc{RAD}}$,
\be\label{first term}
\left(y,z,\frac{\z^2-k^2\xi}{2\xi x^2y}\right),
\ee
cannot be further splitted in its two parts: Each one of the resulting terms in (\ref{first term}), if taken alone asymptotically, leads to 64 decompositions for which the subdominant part either grows or tends to a constant, rather than decays to zero as $t\rightarrow 0$. Of the  remaining 127 decompositions, 64 contain the term (\ref{first term}) while 63  do not contain it in their dominant  $\mathbf{f}^{(0)}_{\,k,\textsc{RAD}}$ part.
Also, 15 of the 127 decompositions of the field do not contain in their dominant part neither the term
(\ref{first term}) nor any of the terms containing the curvature $k$.
It is interesting that of these 15 decompositions, 7 do not contain the linear term $-y/2\xi$ in the dominant part, and among these \emph{there is only one} that eventually leads to a fully acceptable (cf. below)  dominant balance: namely, this is the decomposition that does not contain (in its dominant part) any of the terms containing the parameters $\z,k$ as well as the linear term $-y/2\xi$, but does contain all the other remaining three terms of the vector field (\ref{vf}).
We therefore arrive at the interesting  qualitative conclusion that in the higher order gravity asymptotics towards the initial singularity of the curved, radiation-filled FRW models, neither radiation not curvature play a dominant role!

The rest  126 decompositions fail to lead to an acceptable dominant balance for one of the three different reasons: Either
\begin{enumerate}
\item the dominant exponents in a balance (cf. (\ref{eq:domisol})) cannot be determined as a solution of the dominant system, or
\item the condition for the existence of a acceptable subdominant part \be\label{cond sub}
\lim_{t\rightarrow 0} \frac{\mathbf{f}^{\,(\textrm{sub})}_{\,k,\textsc{RAD}}(\mathbf{a}t^{\mathbf{p}})}{t^{\mathbf{p-1}}}=0, \ee is not satisfied, or
\item they lead to a conflict with the arbitrariness of the position of the singularity (this translates into an invalid condition on the eigenvalues of the so-called Kovalevskaya matrix, cf. \cite{CB} and below).
\end{enumerate}
   We conclude that the only acceptable asymptotic splitting of the vector field  $\mathbf{f}^{(0)}_{\,k,\textsc{RAD}}$ is of the form (\ref{general split}),
$$
\mathbf{f}_{\,k,\textsc{RAD}}=\mathbf{f}^{(0)}_{\,k,\textsc{RAD}} + \mathbf{f}^{\,(\textrm{sub})}_{\,k,\textsc{RAD}},
$$
with with dominant part
\be
\mathbf{f}^{(0)}_{\,k,\textsc{RAD}}(\mathbf{x})=\left(y,z,\frac{3y^3}{2x^2} + \frac{z^2}{2y} -\frac{yz}{x}
\right), \label{eq:f0}
\ee
and subdominant part
\be\label{eq:f01}
\mathbf{f}^{\,(\textrm{sub})}_{\,k,\textsc{RAD}}(\mathbf{x})=
\left(0,0, \frac{\z^2-k^2\xi}{2\xi x^2y} - \frac{y}{2\xi} -\frac{k}{2\xi y} + \frac{ky}{x^2}\right).
\ee
A final comment about the asymptotic splittings of the field equations of the present Section is in order. It is interesting that the dominant part of the vector field given by Eq. (\ref{eq:f0}) is precisely the same as that of the field in the flat, radiation dominated case treated in \cite{K1} (see Eq. (16) in that paper). Their \emph{difference} lies in the subdominant parts of the two cases, the curved one treated here and the flat case in \cite{K1}: Here the subdominant part given by (\ref{eq:f01}) contains precisely the terms of the vector field  $\mathbf{f}^{\,(\textrm{sub})}_{\,0,\textsc{RAD}}$ (radiation and linear terms),  plus the three curvature terms (those with a $k$ in Eq. (\ref{vf}).

We have completed in this Section the first part of our asymptotic analysis via the method of asymptotic splittings, that is we found all possible asymptotic systems on approach to the initial singularity. This process amounts to dropping all terms that are \emph{small}, and replace exact by asymptotic relations (by this we mean using (\ref{basic dom ds}) in conjunction with (\ref{general split}) and (\ref{cond sub}) instead of (\ref{basic dynamical system}) and (\ref{vf})). This first part of the application of the method of asymptotic splittings allows us to conclude that there is essentially a unique such system, and we were able to extract some qualitative results about the behaviour of our basic vector field without actually solving the systems. In the next Section, we shall proceed to study the solutions of our asymptotic system through the processes of \emph{balance}, \emph{subdominance} and \emph{consistency}.

\section{Stability of the asymptotic solutions}
We now look for the possible asymptotic solutions, asymptotic forms of integral curves  of the curvature-radiation field $\mathbf{f}_{\,k,\textsc{RAD}}$, that is we search for the dominant balances determined by the dominant part $\mathbf{f}^{(0)}_{\,k,\textsc{RAD}}$ given by Eq. (\ref{eq:f0}). For this purpose, we substitute in the dominant system $(\dot x,\dot y,\dot z)(t)=\mathbf{f}^{(0)}_{\,k,\textsc{RAD}}$ the forms (\ref{eq:domisol}) and solve the resulting nonlinear algebraic system to determine  the dominant balance $(\mathbf{a},\mathbf{p})$ as an exact, scale invariant solution. This leads to the unique \emph{curvature-radiation} balance $\mathcal{B}_{\,k,\textsc{RAD}}\in\mathbb{C}^3\times\mathbb{Q}^3$,  with
\be\label{basic balance}
\mathcal{B}_{\,k,\textsc{RAD}}=(\mathbf{a},\mathbf{p})= \left(\left(
\t,\frac{\t}{2},-\frac{\t}{4}\right),\:
\left(\frac{1}{2},-\frac{1}{2},-\frac{3}{2}\right)\right),
\ee
where $\t$ is a real, arbitrary constant. In particular, this means that the vector field $\mathbf{f}^{(0)}_{\,k,\textsc{RAD}}$ is \emph{a scale invariant system}, cf.
\cite{CB,go}.

Further, we need to show that   the higher order terms (\ref{eq:f01}) in the basic decomposition of the curvature-radiation field (\ref{general split}) are themselves weight-homogeneous with respect to the curvature-radiation balance (\ref{basic balance}) for this  to be an acceptable one. To prove this, we first split the subdominant part (\ref{eq:f01})  by writing
\be
\mathbf{f}^{\,(\textrm{sub})}_{\,k,\textsc{RAD}}(\mathbf{x}) = \mathbf{f}^{(1)}_{\,k,\textsc{RAD}}(\mathbf{x}) +
\mathbf{f}^{(2)}_{\,k,\textsc{RAD}}(\mathbf{x}) + \mathbf{f}^{(3)}_{\,k,\textsc{RAD}}(\mathbf{x}),
\ee
where
\be
\mathbf{f}^{(1)}_{\,k,\textsc{RAD}}(\mathbf{x})=\left(0,0,\frac{ky}{x^2}\right),\,\,
\mathbf{f}^{(2)}_{\,k,\textsc{RAD}}(\mathbf{x})=\left(0,0,\frac{\z^2-k^2\xi}{2\xi x^2 y}-\frac{y}{2\xi}\right),\,\,
\mathbf{f}^{(3)}_{\,k,\textsc{RAD}}(\mathbf{x})=\left(0,0,-\frac{k}{2\xi y}\right).
\ee
We can then find the required condition that it is indeed subdominant. Using the balance $\mathcal{B}_{\,k,\textsc{RAD}} $ defined by Eq. (\ref{basic balance}), we find that (note that $t^{r-1}=t^{-5/2}$)
\bq
\frac{\mathbf{f}^{(1)}_{\,k,\textsc{RAD}}(\mathbf{a}t^{\mathbf{p}})}{t^{\mathbf{p}-1}}&=&
\mathbf{f}^{(1)}_{\,k,\textsc{RAD}}(\mathbf{a})t,\\
\frac{\mathbf{f}^{(2)}_{\,k,\textsc{RAD}}(\mathbf{a}t^{\mathbf{p}})}{t^{\mathbf{p}-1}}&=&
\mathbf{f}^{(2)}_{\,k,\textsc{RAD}}(\mathbf{a})t^{2},\\
\frac{\mathbf{f}^{(3)}_{\,k,\textsc{RAD}}(\mathbf{a}t^{\mathbf{p}})}{t^{\mathbf{p}-1}}&=&
\mathbf{f}^{(3)}_{\,k,\textsc{RAD}}(\mathbf{a})t^{3}.
\eq
Hence, taking the limit as $t\rightarrow 0$, we see that these fractions go to zero asymptotically provided that the forms $\mathbf{f}^{(i)}_{\,k,\textsc{RAD}}(\mathbf{a}), i=1,2,3,$ are  all different from zero. This happens only when
$\xi\neq 0$, that is for all cases except when $3\b +\g +\delta=0$. We conclude that this result is true in all higher order gravity theories \emph{except} perhaps the so-called conformally invariant Bach-Weyl gravity\footnote{Apparently, this case needs a separate treatment altogether.}, cf. \cite{bach+weyl}, that emerges when $\delta =0$.
Since the \emph{subdominant exponents}
\be\label{sub exps}
q^{(0)}=0\ <\ q^{(1)}=1\ <\ q^{(2)}=2\ <\ q^{(3)}=3,
\ee
are ordered, we conclude that the subdominant part (\ref{eq:f01}) is weight-homogeneous as promised.

Let us now move on to the last part of our asymptotic investigations. For our asymptotic solutions to be valid, we further need to check their \emph{consistency} with the overall approximation scheme we are using. This means that we need to construct a series representation of the asymptotic solutions valid locally around the initial singularity, so that it is dominated by the dominant balance solutions we have built so far. The degree of generality of these final series solutions will of course depend on the number of arbitrary constants in them. We expect such constants to appear in certain places in the final formal developments we are after and, as an intermediate step in the construction of the series expansions,  we now calculate the precise positions where such constants would appear.

As explained in \cite{CB}, the arbitrary constants of any
(particular or general) solution first appear in those terms in the
asymptotic series solution whose coefficients $\mathbf{c}_{i}$ have
indices $i=\varrho s$, where $\varrho$ is a non-negative
$\mathcal{K}$-exponent and $s$ denotes the least common multiple of the denominators of the set of all subdominant exponents (\ref{sub exps}) and those of all the $\mathcal{K}$-exponents with positive real parts (in our case, $s=2$). These exponents are numbers belonging to the spectrum of an important matrix in asymptotic analysis, the so-called Kovalevskaya matrix given by
\be
\mathcal{K}=D\,\mathbf{f}^{(0)}_{\,k,\textsc{RAD}}(\mathbf{a})-\textrm{diag}(\mathbf{p}).
\ee
Hence, the $\mathcal{K}$-exponents depend of the dominant part of the vector field as well as the dominant balance. In our case, the Kovalevskaya matrix is
\be
\mathcal{K}_{\,k,\textsc{RAD}}=\left(
                     \begin{array}{ccc}
                       -1/2& 1  & 0\\
                       0 & 1/2&1 \\
                       -1/2&5/4&1/2
                     \end{array}
                   \right),
\ee
with spectrum
\be
\textrm{spec}(\mathcal{K}_{\,k,\textsc{RAD}})=\{-1,0,3/2\},
\ee
and corresponding eigenvectors
\be \{(4,-2,3),(4,2,-1),(1,2,2)\}.  \ee
The number of non-negative
$\mathcal{K}$-exponents equals the number of arbitrary
constants that appear in the series expansions. There is always the
$-1$ exponent that corresponds to an arbitrary constant, the
position of the singularity, and because the $\textrm{spec}(\mathcal{K}_{\,k,\textsc{RAD}})$ in our case possesses two non-negative eigenvalues, the balance $\mathcal{B}_{\,k,\textsc{RAD}}$ indeed corresponds to the dominant behaviour of a \emph{general} solution having the form of a formal series and valid locally around the initial singularity. To find it, we substitute the \emph{Puiseux series expansions}
\begin{equation} \label{eq:series}
x(t) = \sum_{i=0}^{\infty} c_{1i} t^{\frac{i}{2}+\frac{1}{2}}, \:\:\:\:\:
y(t) = \sum_{i=0}^{\infty} c_{2i}  t^{\frac{i}{2}-\frac{1}{2}},\:\:\:\:\:
z(t) = \sum_{i=0}^{\infty} c_{3i} t^{\frac{i}{2}-\frac{3}{2}} ,\:\:\:\:\:
\end{equation}
where  $c_{10}=\t ,c_{20}=\t /2,c_{30}=-\t
/4$, in the original system (\ref{eq:ds}) and we are led to various recursion relations that determine the unknowns $c_{1i}, c_{2i}, c_{3i}$ term by term. Further algebraic manipulations lead to the final series representation of the solution in the form:
\be
x(t) = \t \:\:t^{1/2} - \frac{k}{2\t}\:\:t^{3/2} +
c_{13} \:\: t^{2} + \displaystyle
\left(\frac{4\z^2-\t^4}{12\xi\t^3}-\frac{k^2}{8\t^3}\right)\:\:t^{5/2} + \cdots .
\label{eq:gensol}
\ee
The corresponding series expansions for $y(t)$ and $z(t)$ are
given by  the first and second time derivatives of the above
expression respectively.

As a final test for admission of this solution, we use Fredholm's
alternative to be satisfied by any admissible solution. This leads to
a \emph{compatibility condition} for the positive
eigenvalue 3/2 and the associated eigenvector: This condition has the form
\be
v^{\top}\cdot\left(\mathcal{K}-\frac{j}{s}I\right)\mathbf{c}_j=0,
\ee
where $I$ denotes the identity matrix, and we have to satisfy
this at the $j=3$ level. This gives the following orthogonality constraint,
\be (1,2,2) \cdot \left( \begin{array}{l}
                       -2c_{13}+c_{23}  \\
                       -c_{23}+c_{33}  \\
                       -\frac{1}{2}c_{13}+\frac{5}{4}c_{23}-c_{33}
                     \end{array}
              \right) = 0.
\label{eq:cc}
\ee
It is immediate to check that this is indeed  satisfied, thus leading to the conclusion that (\ref{eq:gensol}) corresponds indeed to a valid asymptotic solution around the singularity.  We also see that setting $k=0$ in the series expansion we are led exactly to the form found for the flat, radiation case, cf. Eq. (21)\footnote{In that reference, $12\xi\t^3$ was mistakenly written as $24\xi\t^3$.} of \cite{K1}.

Our series (\ref{eq:gensol}) has three arbitrary constants, $\t,
c_{13}$ and a third one corresponding to the arbitrary position of
the singularity (taken here to be zero without loss of generality), and is therefore a local expansion of the
\emph{general} solution around the initial singularity. Since
the leading order coefficients are real, by a theorem
of Goriely and Hyde, cf.  \cite{go}, we conclude that there is an open set of
 initial conditions  for which the general solution blows up at the
finite time (initial) singularity at $t=0$.  This proves the
stability of our solutions in the neighborhood of the singularity.

\section{Conclusions}
In this paper we have considered the behaviour of the flat, radiation-filled isotropic and homogeneous cosmological models in the general quadratic gravity theory near the initial singularity. We have shown that asymptotically the effects of curvature and radiation on the global evolution become negligible and the initial state of these universes is basically described by the flat, vacuum $t^{1/2}$ solution of these theories (with the possible exception of the solutions in the conformally invariant Bach-Weyl theory - this case  needs to be analyzed separately). In particular, all curved radiation solutions tend to this solution asymptotically as we approach the initial state in these theories.

There are precisely 255 ways that the basic curvature-radiation vector field of this problem can split and each one of these needs to be checked for admissibility for it could in principle contain various dominant asymptotic solutions. However, taking into account various asymptotic conditions that have to hold in order to have admissible solutions, we are left with only one possible asymptotic decomposition of the vector field near the initial state. Further, it turns out that this asymptotic form admits only one dominant balance, exact solution of the scale invariant system, and this is precisely the form that behaves as $t^{1/2}$ near the singularity. Using various asymptotic and geometric arguments, we were able to built a solution of the field equations in the form of a Puiseux formal series expansion compatible with all other constraints, dominated asymptotically by the $t^{1/2}$ solution and having the correct number of arbitrary constants that makes it a general solution. In this way, we can conclude that this exact solution is an attractor of all homogeneous and isotropic radiation solutions of the theory thus proving stability against such `perturbations'.

The impressive restrictions placed by the higher order field equations on the structure of the possible initial cosmological states of the theory imply that the initial state of these radiation universes probably resembles that of a vacuum, flat  model  since, as we showed here, the only possible mode of approach to the singularity is one  in which both the curvature \emph{and} radiation modes enter asymptotically only in the subdominant part of the vector field. In fact, an investigation of the asymptotic stability of the flat, vacuum initial state in these theories with respect curved FRW perturbations is currently under scrutiny, cf. \cite{kolionis2}. If both states prove to be stable ones in a way along the lines of the present paper, whether or not the radiation or the vacuum initial states are in addition generic ones in the space of solutions of the higher order gravity equations is currently an open problem.

\section*{Acknowledgements}
We are grateful to Ilya L. Shapiro for an illuminating remark.

\end{document}